\documentstyle[11pt,paspconf,epsf]{article}

\begin{document}

\title{The Kinematics of Galactic Stellar Disks}

\author{Michael R. Merrifield}
\affil{Department of Physics \& Astronomy, University of Southampton,
Highfield, Southampton, SO17 1BJ, England}

\author{Konrad Kuijken}
\affil{Kapteyn Instituut, PO Box 800. 9700 AV Groningen, The Netherlands}

\begin{abstract}
The disks of galaxies are primarily stellar systems, and fundamentally
dynamical entities.  Thus, to fully understand galactic disks, we must
study their stellar kinematics as well as their morphologies.
Observational techniques have now advanced to a point where quite
detailed stellar-kinematic information can be extracted from spectral
observations.  This review presents three illustrative examples of
analyses that make use of such information to study the formation and
evolution of these systems: the derivation of the pattern speed of the
bar in NGC~936; the calculation of the complete velocity ellipsoid of
random motions in NGC~488; and the strange phenomenon of
counter-rotation seen in NGC~3593.

\end{abstract}

% Keywords should be included, but they are not printed in the hardcopy.


% That's it for the front matter.  On to the main body of the paper.
% We'll only put in tutorial remarks at the beginning of each section
% so you can see entire sections together.

\section{Introduction}

Disk galaxies, with their pleasing symmetry and intricate spiral
structure, are amongst the most beautiful of astronomical objects.
They owe this beauty to two basic factors.  First, and fairly
obviously, they are visible to us because they are made up of stars
emitting at optical wavelengths.  Second, they are fundamentally
dynamical entities.  As first recognized by Immanuel Kant, their basic
disk shapes arise from the ordered angular momenta of their
constituent stars.  Even the finer detail of spiral arms, bars, rings,
and so on, can only be understood in terms of the collective motions
of stars.  

The stellar motions also provide a key to studying the formation and
evolution of galaxies.  The two-body relaxation time for close
encounters between stars to significantly re-arrange their orbits is
$\sim 10^{16}\,{\rm years}$.  Thus, the only processes that can affect
the arrangement of stellar orbits are the original formation of the
galaxy, and any large-scale collective phenomena such as those
produced by dynamical instabilities or interactions with other
galaxies.  By studying the arrangement of orbits in a galaxy, we are
tapping directly into the archaeological record of these processes.

Finally, it should be remembered that the observed stellar component
is the tip of the galactic iceberg: the total mass of galaxies seems
to be dominated by non-luminous ``dark matter.''  Since the
gravitational potential dictating the stars' orbits is generated by
all the mass in the galaxy, the arrangement of stellar orbits provides
one of the few direct mechanisms by which we can ``see'' this
invisible component.  The simplest analyses involve deriving the
over-all distribution of mass in a galaxy from the circular motions of
stars around the disk at different radii (e.g. Kent 1987).  However --
as we shall see in Section~\ref{TWsec} -- it is also possible to draw
subtler inferences about the distribution of dark matter in disk
galaxies from the systems' stellar-kinematic properties.

\begin{figure}
\plotone{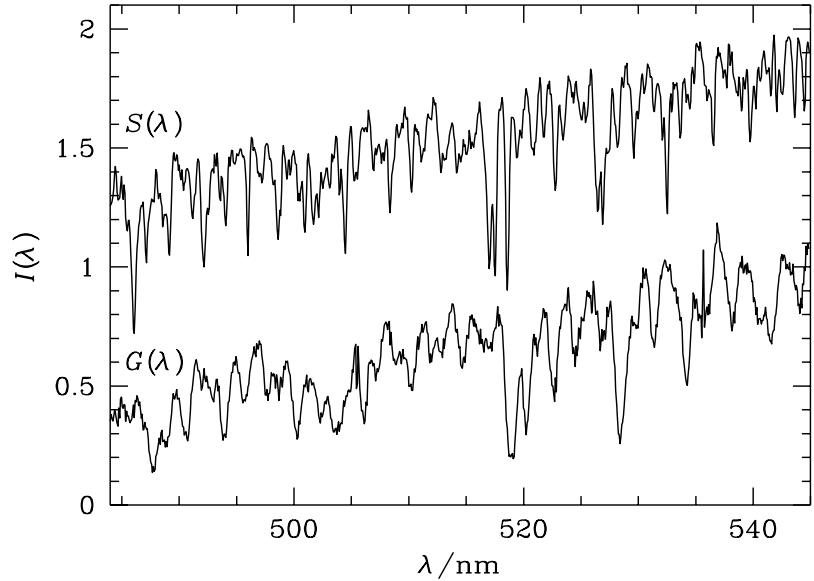}
\caption{Intensity as a function of wavelength for a K0 giant star 
[$S(\lambda)$] and a small part of a galaxy [$G(\lambda)$].  The
strong absorption lines at $\sim517\,{\rm nm}$ are the Mg~b feature,
and most of the other lines are due to Fe.  Note the Doppler
broadening and shift in the galaxy spectrum.  [Reproduced from Binney
\& Merrifield (1998).]}
\label{specfig}
\end{figure}

Thus, if we want to understand the formation and evolution of disk
galaxies, and probe the underlying distribution of mass in such
systems, we need to investigate the motions of their constituent
stars.  The primary technique for studying such motions is through the
Doppler shifts of the absorption lines in the stars' spectra.
Unfortunately, we cannot observe individual stars in any but the
closest of galaxies; typically, a spectrum of the smallest resolvable
element of a galaxy will contain the light from hundreds of thousands
of stars.  However, there is still plenty of dynamical information in
the composite galaxy spectrum: the spectrum of each contributing star
will be Doppler shifted by a slightly different amount, so the
absorption lines in the integrated spectrum will be significantly
broader than those in a stellar spectrum (see Figure~\ref{specfig}).
In fact, if we assume that all the stars in the galaxy have identical
spectra, then the observed galaxy spectrum will be simply a
convolution of the stellar spectrum with a broadening function that
represents the distribution of line-of-sight velocities of all the
stars that contribute to the light.  Thus, if we obtain both a galaxy
and a stellar spectrum, we can derive the line-of-sight-velocity
distribution of the stars in the galaxy spectrum through a process of
deconvolution.

In practice, extracting this dynamical information is not a trivial
process.  First, galactic disks are intrinsically faint objects: in
their most luminous central regions, they are only comparable to the
brightness of the night sky.  Thus, it is difficult and time consuming
to obtain galaxy spectra at even moderate signal-to-noise ratios.
However, the development of high-throughput spectrographs and
efficient detectors has made it possible to dramatically improve the
quality of galactic spectra that can be attained, and the arrival of
8m-class telescopes will make such data routinely available.  

The second problem is that the deconvolution process greatly amplifies
any noise in the data.  Hence, even a galaxy spectrum of high quality
will only yield a very poor estimate for the distribution of stellar
velocities unless some care is taken.  The traditional solution to
this problem has been to assume that the stellar line-of-sight
velocity distribution has a particularly simple form, which is usually
taken to be a Gaussian (e.g.\ Sargent et al.\ 1977, Tonry \& Davis
1979).  The kinematics of the galaxy are then quantified by finding
the best-fit Gaussian such that, when the stellar spectrum is
convolved with this distribution, the resulting broadened spectrum
most closely matches the galaxy data.  The mean of the Gaussian then
provides a measure of the mean streaming motion of the stars, while
its dispersion quantifies the stellar random motions.  More recently,
algorithms have been developed that allow a more general deconvolution
of spectral data without uncontrollable amplification of the noise
(e.g.\ Gerhard 1993, van der Marel \& Franx 1993, Kuijken \&
Merrifield 1993).  Using such techniques, it is now possible to go
beyond the traditional crude measures of stellar kinematics to
estimate the complete line-of-sight velocity distribution
of all the stars that contribute to a galaxy spectrum.

In this review, we present the results of three recent projects that
investigate the dynamics of stellar disks.  These examples are not
intended to provide a comprehensive overview of current work on
stellar disk dynamics; rather, they have been chosen simply to
illustrate the variety of information that can be gleaned from such
studies.  Section~\ref{TWsec} describes the calculation of the pattern
speed of the bar in NGC~936 from the stellar mean streaming motions in
this galaxy.  As we discuss, in addition to being a fundamental
property of barred galaxies, the derived pattern speed also has
implications for the distribution of dark matter in galaxies.  The
second example, presented in Section~\ref{heatsec}, shows how the
three-dimensional distribution of random motions within the disk
galaxy NGC~488 can be derived from the broadening of its spectral
lines, and how the balance between the velocity dispersions in
different directions can be used to identify the ``heating'' process
responsible for producing these random motions.  Finally,
Section~\ref{CRsec} illustrates the strange phenomenon of
counter-rotation detected in a few stellar disks by presenting the
complete line-of-sight velocity distribution for the stars in the
edge-on disk galaxy NGC~3593.

\section{The Pattern Speed of NGC~936} \label{TWsec}

\begin{figure}
\plottwo{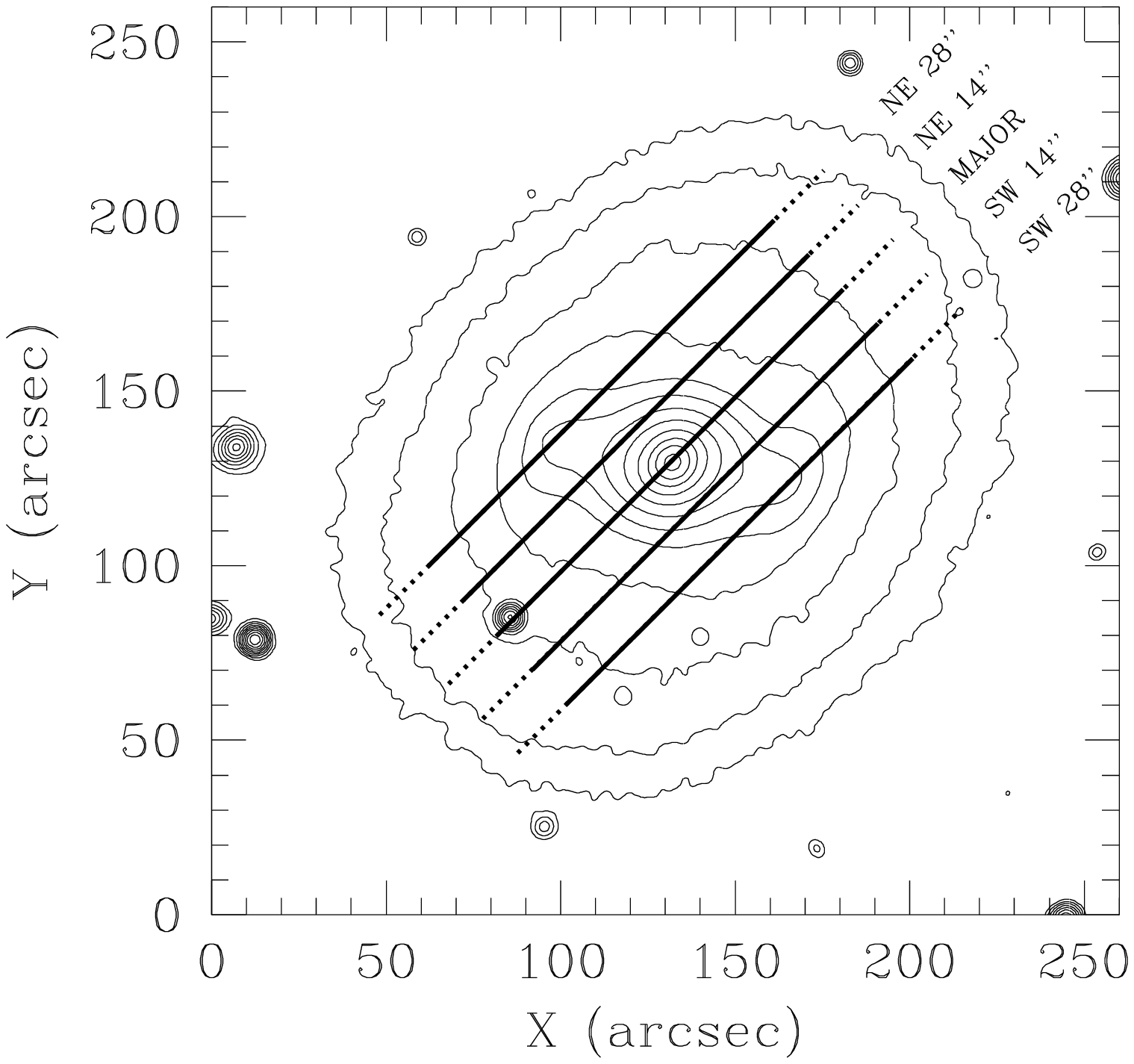}{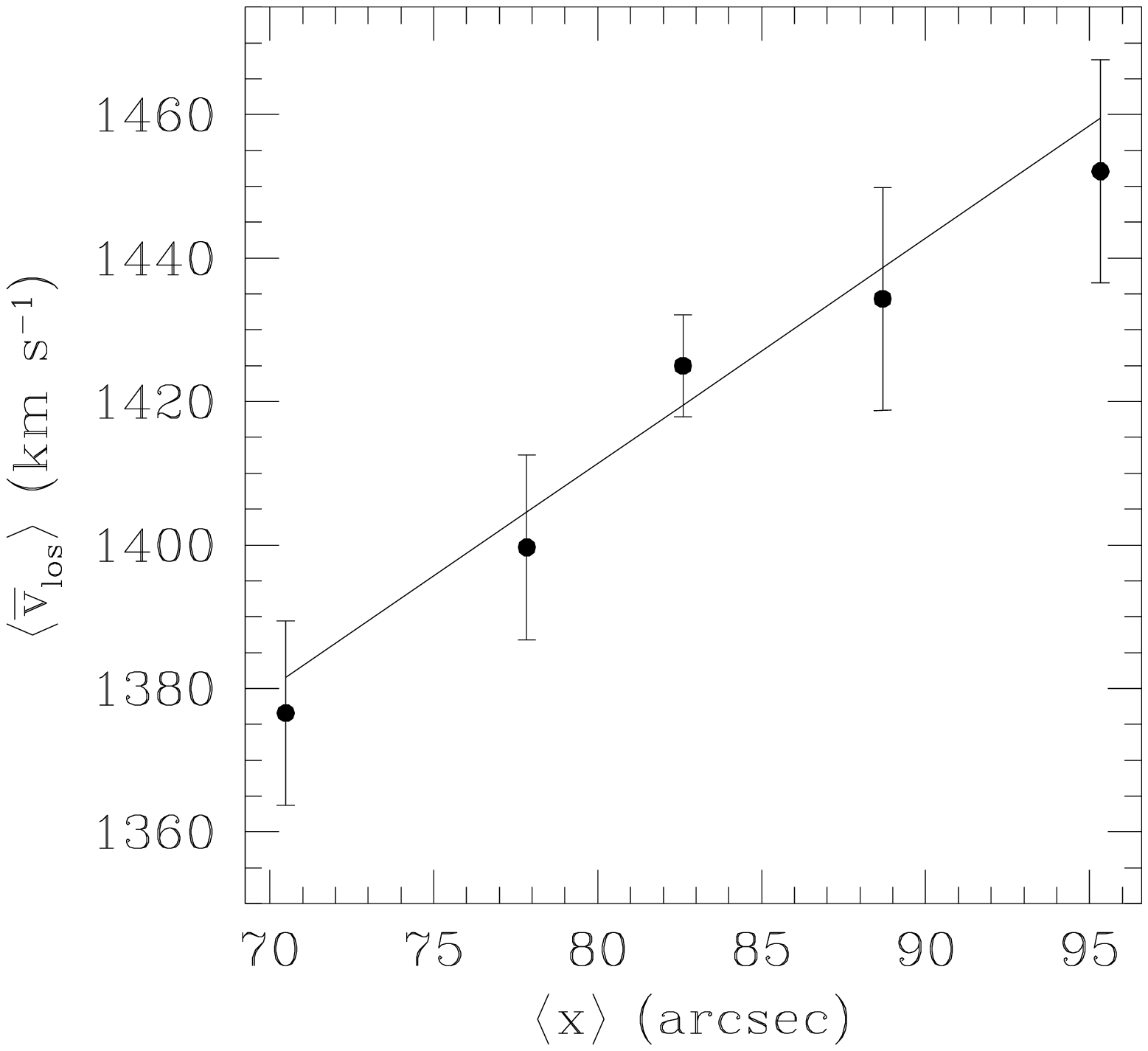}
\caption{Calculation of the pattern speed of NGC~936.  The left panel
shows a contour plot of an I-band image of the galaxy overlayed by the
positions at which the spectra were obtained.  the right panel shows
the resulting plot of $\langle \bar v_{\rm los} \rangle$ against
$\langle x \rangle$, whose slope is $\Omega_p \sin i$.  [Adapted from
Merrifield \& Kuijken (1995).]}\label{TWfig}
\end{figure}

Approximately a third of disk galaxies contain a central rectangular
enhancement in their stellar densities (see Figure~\ref{TWfig}, left
panel).  It has long been recognized that such stellar bars are the
product of a dynamical instability in self-gravitating disks (e.g.\
Miller, Prendergast \& Quirk 1970).  Over time, a bar will rotate
about the center of its galaxy, but it will not generally do so at the
same rate as the stars in the galaxy, so one cannot associate
particular stars with permanent membership of the bar; rather, one
should think of the bar as a density wave propagating through the
stars of the disk.  The absence of a direct link between the rate at
which the bar rotates (its ``pattern speed'') and the rate at which
stars rotate means that the pattern speed cannot be obtained trivially
from the streaming motions of the stars.  However, Tremaine \&
Weinberg (1986) showed that these two quantities can be related by
invoking the continuity equation (which merely imposes the constraint
that stars should not disappear or appear as they orbit around
galaxy).  They thus showed that
\begin{equation}\label{TWeq}
\Omega_p \sin i \int_{-\infty}^\infty I x\, {\rm d}x =
\int_{-\infty}^\infty I \bar v_{\rm los}\,{\rm d}x,
\end{equation}
where $\Omega_p$ is the pattern speed of the bar (in radians per year,
or, more commonly, ${\rm km}\,{\rm s}^{-1}\,{\rm kpc}^{-1}$), $i$ is
the inclination of the galaxy, $\bar v_{\rm los}$ is the mean
streaming velocity of the stars at each point in the disk, $I$ is the
surface brightness at each point, and the integrals are carried out
along any line parallel to the galaxy's major axis.  Hence, if we
denote the intensity-weighted averages of $x$ and $\bar v_{\rm los}$
along lines parallel to the major axis as $\langle x \rangle$ and
$\langle \bar v_{\rm los} \rangle$ respectively, we find that
$\Omega_p \sin i \langle x \rangle = \langle \bar v_{\rm los}
\rangle$.  If we calculate $\langle x \rangle$ and $\langle \bar
v_{\rm los} \rangle$ for a series of lines parallel to the galaxy's
major axis, then a plot of one against the other should yield a
straight line of slope $\Omega_p \sin i$.  Using the apparent axial
ratio of the galaxy disk at large radii (where it is assumed to be
intrinsically round), we can estimate $i$, and hence solve for
$\Omega_p$.

Although elegantly simple in principle, the practical implementation
of this Tremaine--Weinberg algorithm has proved difficult.  For a
start, if a galaxy contains significant regions of on-going star
formation, then the continuity equation upon which the method relies
is not valid, as the quantity of stars (and, more importantly, the
amount of light they produce) is not conserved as they travel around
the galaxy.  Second, the presence of patchy extinction due to dust in
the galaxy will alter the amount of light that we receive from a star
at different points around its orbit, which also invalidates the use
of the continuity equation.  Finally, the contributions to the
integrals in equation~(\ref{TWeq}) from the range $-\infty < x < 0$
are almost canceled by the contributions from $0 < x < \infty$, so
$\Omega_p$ is given by the ratio of two integrals that are both close
to zero.  Thus, any noise in the data will be strongly amplified,
resulting in large errors in the derived value of $\Omega_p$.  Until
recently, it has not been possible to reduce the noise in the value of
$\langle \bar v_{\rm los} \rangle$ derived from spectra to a point
where a meaningful value for $\Omega_p$ could be obtained.

The first successful implementation of the Tremaine--Weinberg method
was to observations of the SB0 galaxy NGC~936 (Merrifield \& Kuijken
1995).  An S0 galaxy is well suited to this analysis, as it contains
little obscuring dust and no significant star formation.  NGC~936 is
also ideally oriented, since having the bar at $45\,{\rm degrees}$ to
the major axis (see Figure~\ref{TWfig}) maximizes the signal in the
integrals in equation~(\ref{TWeq}).  Using the Multiple-Mirror
Telescope, we obtained five sets of spectra with the spectrograph slit
oriented parallel to the major axis, as shown in Figure~\ref{TWfig}.
From these data, we obtained the over-all brightness of the galaxy as
a function of position along each slit, $I(x)$, from which $\langle x
\rangle$ was calculated.  We also calculated the line-of-sight
velocity distribution of all the stars whose light was admitted by the
slit by adding together all the spectra obtained through the slit, and
using a deconvolution technique to extract the velocity distribution
from the Doppler broadening of the composite spectrum.  The quantity
$\langle \bar v_{\rm los} \rangle$ is then simply the mean velocity of
this integrated distribution.  As Figure~\ref{TWfig} shows, a plot of
$\langle x \rangle$ against $\langle \bar v_{\rm los} \rangle$ for the
various slit positions does, indeed, yield a straight line.  Combining
this slope with the inclination derived from the galaxy's photometry
($i = 41\,{\rm deg}$), we find $\Omega_p = 60 \pm 14\,{\rm km}\,{\rm
s}^{-1}\,{\rm kpc}^{-1}$.

The success of this calculation confirms the basic picture of a bar as
a rotating pattern with a well-defined pattern speed.  Indirectly, it
also tells us something about the environment in which the bar has
formed.  In particular, it is instructive to calculate the
``co-rotation radius'' of the bar, which is the radius at which the
stars travel around the galaxy at the same angular speed as the bar
pattern.  Numerical simulations have shown that when a bar forms, the
pattern rotates at such a rate that the co-rotation radius lies just
outside the end of the bar.  The subsequent evolution of the pattern
speed depends on the bar's environment.  In isolation, the bar will
continue to rotate at this high rate.  If, however, the bar is
embedded in a massive halo of material, then interactions between the
bar and the halo -- a form of dynamical friction -- will cause the bar
to transfer angular momentum to the halo and slow down (Debattista \&
Sellwood 1998).  If one were to calculate the co-rotation radius of
such a decelerated bar, one would find that it lay well beyond the end
of the bar.

When one combines the derived value of $\Omega_p$ with the rate of
stellar rotation as a function of radius for NGC~936, one finds that
the co-rotation radius lies just beyond the end of the photometric
bar.  Thus, unless we have caught this bar exceptionally soon after it
formed, we can conclude that it cannot be embedded in a dense massive
halo.

This conclusion is intriguing because recent cosmological simulations
have suggested that if the Universe is dominated by cold dark matter,
then the galaxy-scale dark matter halos that condense from this
material should have a common centrally-concentrated mass distribution
(Navarro, Frenk \& White 1997).  A disk galaxy forming in such a
centrally-concentrated halo would contain a great deal of dark matter
at small radii, and so dynamical friction would rapidly decelerate any
bar that it might contain.  If NGC~936's rapidly-rotating bar turns
out to be a common phenomenon, the cosmologists will have to think
again.

\section{Disk Heating in NGC~488} \label{heatsec}

As a second example of what can be learned about galactic evolution
from stellar kinematics, we turn to the phenomenon of disk heating.
Although galactic disks are highly-flattened structures, they are not
infinitely thin; an edge-on disk galaxy typically has a thickness of
some hundreds of parsecs (e.g.\ van der Kruit \& Searle 1981).  This
thickness implies that the constituent stars must possess random
motions perpendicular to the plane of the galaxy, which result in the
observed vertical excursions.  Spectral observations of face-on
galaxies (e.g.\ van der Kruit \& Freeman 1986) confirm the presence of
such random motions, as the absorption lines in such spectra are
typically Doppler broadened with dispersions of a few tens of
kilometers per second.  The amplitude of these random motions
decreases steadily with radius, reflecting the weakening gravitational
pull toward the plane of the galaxy as the density of the disk
declines.

Kinematic observations of edge-on galaxies show that there are similar
random motions in the radial and tangential directions within disk
galaxies.  The situation here is a little more complicated, because
the mean streaming velocities of the stars dominate the observable
kinematics of edge-on galaxies, and these streaming motions have
different line-of-sight components at different points on any line
through the galaxy, which also contributes to the total broadening of
the spectral lines.  Further, the radial and tangential random motions
contribute varying amounts to the line-of-sight velocity dispersion at
different points along the line of sight.  However, once these
complexities have been accounted for, we arrive at the same basic
picture of radial and tangential random motions with dispersions of a
few tens of kilometers per second, which decline with distance from
the galactic center (Bottema 1993).

Stars are born in gas clouds, which -- due to their collisional nature
-- follow orbits that are close to circular and lie in the plane of
the galaxy.  Thus, the random motions that we see in the mature
stellar population must have been acquired through some subsequent
``heating'' process.  Two main candidates have been advanced for
generating the random motions.  The first possibility involves heating
by giant molecular clouds.  As a star orbits around its galaxy, it
will occasionally pass sufficiently close to a massive molecular cloud
for it to be gravitationally scattered.  Since such a scattering will
drive the star and cloud toward energy equipartition, the star will on
average be accelerated, increasing its random motions with each such
encounter.  The second candidate for heating also involves
gravitational scattering, but in this case the scattering mass is the
density enhancement associated with a spiral arm.  Each time that an
orbiting star travels through a spiral arm, it will interact with the
high density there, and increase the random component of its motion.

How, then, do we distinguish between these two possible candidates for
the heating mechanism?  A diagnostic was suggested by Jenkins \&
Binney (1990), who calculated the efficiency with which these
mechanisms scatter the stars in different directions.  They showed
that if heating by molecular clouds were the dominant process, then
the amplitudes of the random motions perpendicular to the plane of a
galaxy and those in the radial direction should be related, with
dispersions in the ratio $\sigma_z/\sigma_R \sim 0.75$.  If, on the
other hand, density waves provide the dominant scattering mechanism,
then $\sigma_z/\sigma_R \sim 0.5$.  The basic reason for this
difference is that a star has natural frequencies with which it
oscillates about its original circular orbit.  In the case of radial
motions, this natural frequency amounts to $\sim 2$ oscillations per
complete orbit; the oscillation frequency perpendicular to the plane
is much higher.  A two-armed spiral will give the orbiting star a kick
twice per orbit, which is close to its natural radial frequency, so
the amplitude of motions in this direction will grow very efficiently.
Vertical motions, with their higher natural frequency, do not enjoy
this resonant status, so $\sigma_z/\sigma_R$ will be relatively low
where spiral density waves are responsible for the heating.

Thus, if we can measure the ratio $\sigma_z/\sigma_R$ in a galaxy, we
can straightforwardly determine which heating mechanism is responsible
for the random motions.  There is, of course, a catch: $\sigma_z$ has
only been determined for face-on galaxies, while $\sigma_R$ has only
been measured in inclined systems, so it has not been possible to
determine this ratio for any single galaxy.  The one exception to this
problem is the Milky Way, where combining proper motions of nearby
stars with their line-of-sight velocities enables all three components
of the velocity distribution in the solar neighborhood to be
calculated.  In this case, we find $\sigma_z/\sigma_R \sim 0.5$,
implying that density waves provide the dominant heating mechanism.

With only one datum, it is difficult to draw any general conclusions.
Fortunately, there is a way to measure $\sigma_z/\sigma_R$ in other
galaxies.  If, instead of picking an edge-on or a face-on galaxy, one
looks at an intermediate-inclination system, then it is possible to
determine all three components of the velocity dispersion,
$\{\sigma_R, \sigma_\phi, \sigma_z\}$.  Specifically, the
line-of-sight velocity dispersion along the major axis of a galaxy at
an inclination angle $i$ is given by $\sigma_{\rm maj}^2(R) =
\sigma_\phi^2\sin^2 i + \sigma_z^2 \cos^2 i$, while that along the
minor axis is given by $\sigma_{\rm min}^2(R) = \sigma_R^2\sin^2 i +
\sigma_z^2 \cos^2 i$.  We thus have two independent observable
constraints on the three unknown components of the velocity
dispersion.  To close this system of equations, we need to turn to
dynamical theory: since the non-circular orbit of an individual star
results in both radial and tangential random motions at different
points around the orbit, $\sigma_R$ and $\sigma_\phi$ are not
independent quantities.  In fact, in the epicycle approximation, they
are related by the simple formula,
\begin{equation}\label{epicyceq}
{\sigma_\phi^2 \over \sigma_R^2} = {1 \over 2}\left(1 + {\partial\ln
v_c \over \partial\ln R}\right),
\end{equation}
where $v_c(R)$ is the circular rotation speed (Binney \& Tremaine
1987).  By combining equation~(\ref{epicyceq}) with the observed values
of $\sigma_{\rm maj}(R)$ and $\sigma_{\rm min}(R)$, we can solve for
all three components of the galaxy's intrinsic velocity dispersion.

\begin{figure}
\plottwo{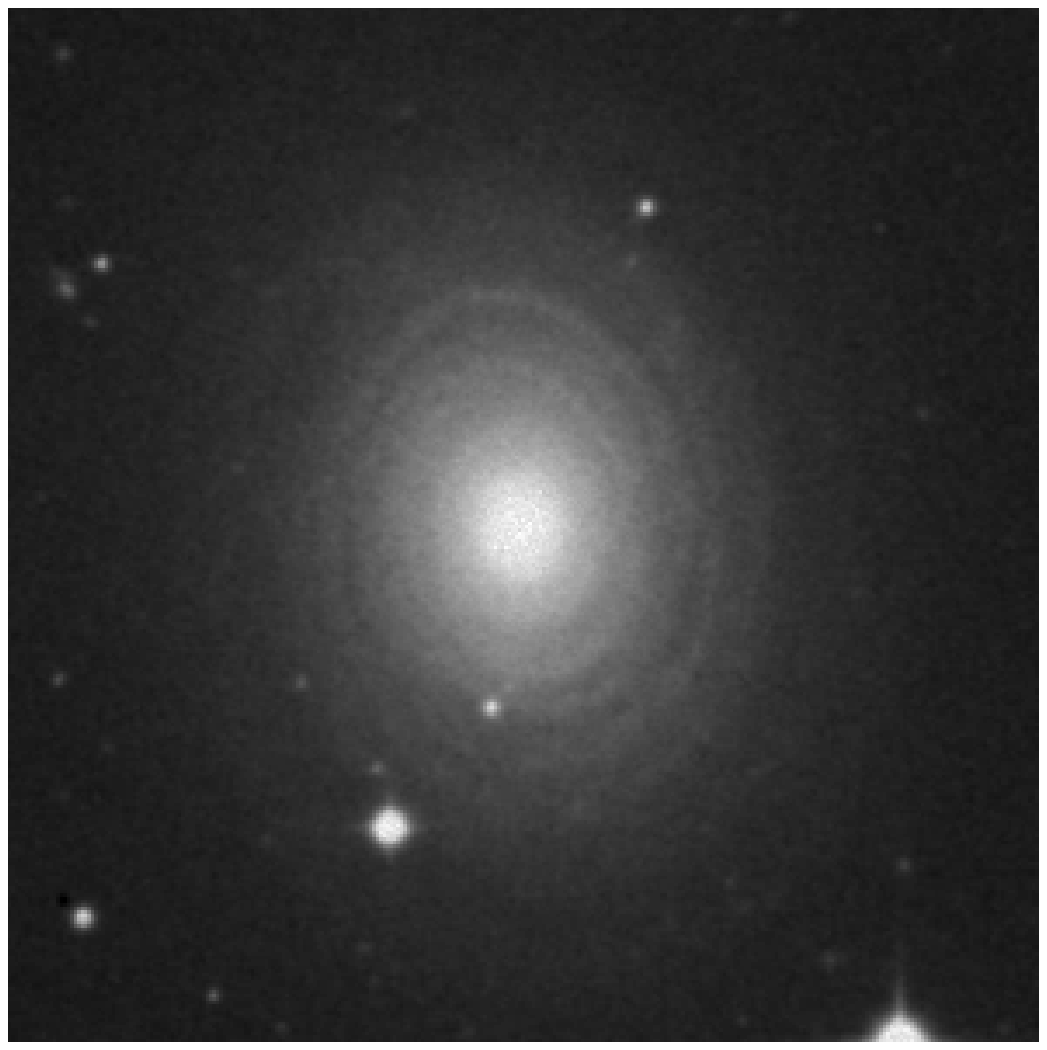}{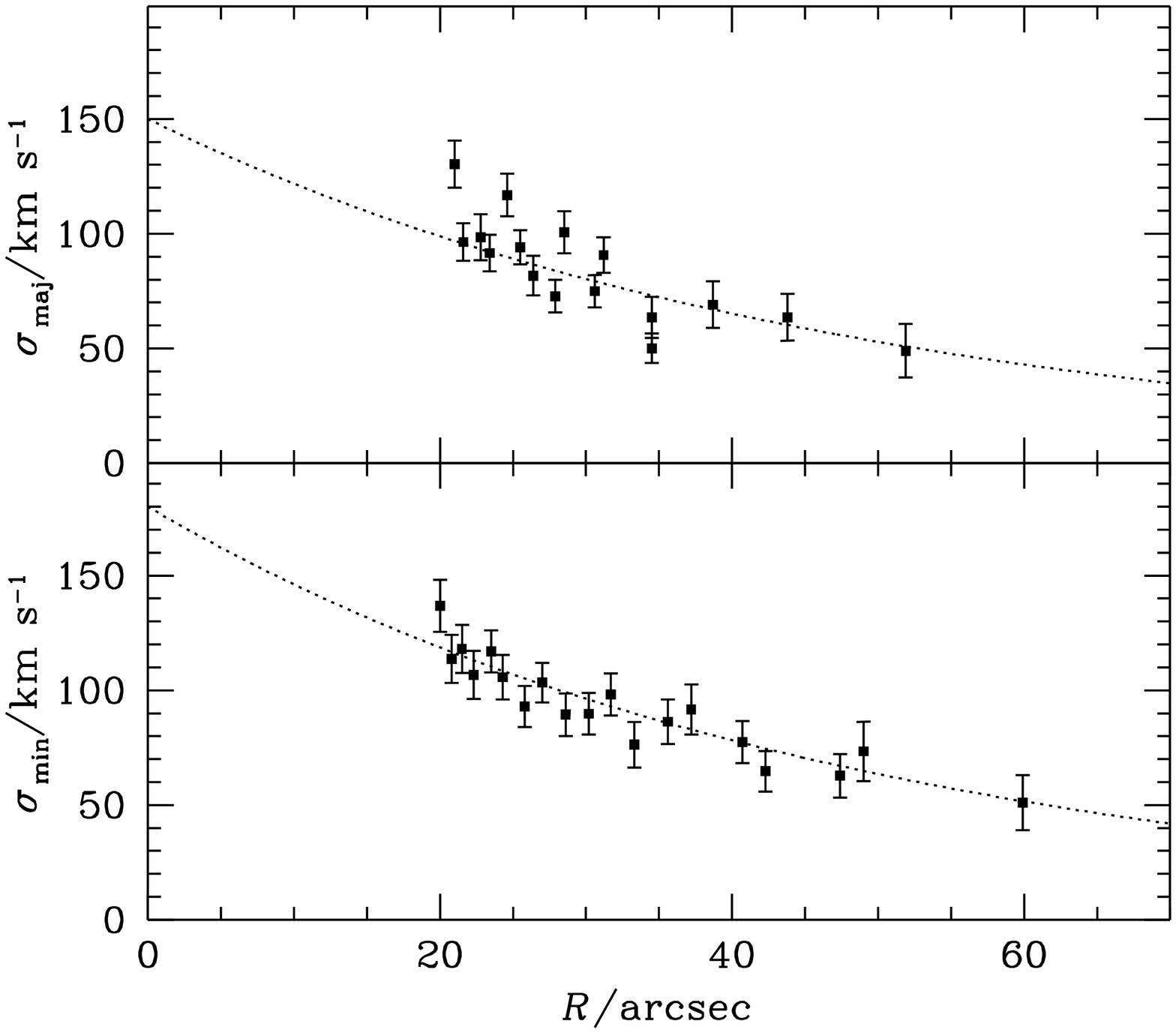}
\caption{Left panel: image of the moderately-inclined Sa galaxy
NGC~488.  Right panel: the line-of-sight velocity dispersion measured
along its major and minor axes.  The dotted line shows the model fit
to these dispersion profiles.  [Adapted from Gerssen, Kuijken \&
Merrifield (1997).]} \label{heatfig}
\end{figure}

In order to apply this approach, we used the William Herschel
Telescope to obtain spectra of the major and minor axes of the
moderately-inclined Sa galaxy NGC~488 (Gerssen, Kuijken \& Merrifield
1997).  Figure~\ref{heatfig} shows the line-of-sight velocity
dispersions derived from these data.  The mean streaming of the stars
and emission-line gas provided the values of $v_c(R)$ for
equation~(\ref{epicyceq}).  Combining this equation with the two
dispersion profiles, we were able to show that the average velocity
dispersions in the three directions lie in the approximate ratio
$\sigma_R:\sigma_\phi:\sigma_z \sim 1:0.7:0.7$.

Unlike the Milky Way, The relatively high value of $\sigma_z/\sigma_R$
for NGC~488 seems to favor molecular clouds as the heating mechanism
in this galaxy.  With hindsight, this difference is not surprising: as
befits its early type, NGC~488 does not contain strong spiral
structure (see Figure~\ref{heatfig}), so it is unlikely that heating
by spiral density waves will be important in this system.  The Milky
Way, on the other hand, is a later-type Sbc galaxy (de Vaucouleurs \&
Pence 1978), where spiral structure will be more significant,
explaining the density waves' leading role in heating the stellar
population.  Thus, we are now in a position to begin to answer the
question of what the dominant stellar heating mechanism is in disk
galaxies: it appears to vary from system to system.

\section{Counter-Rotating Stars in NGC~3593} \label{CRsec}

\begin{figure}
\plottwo{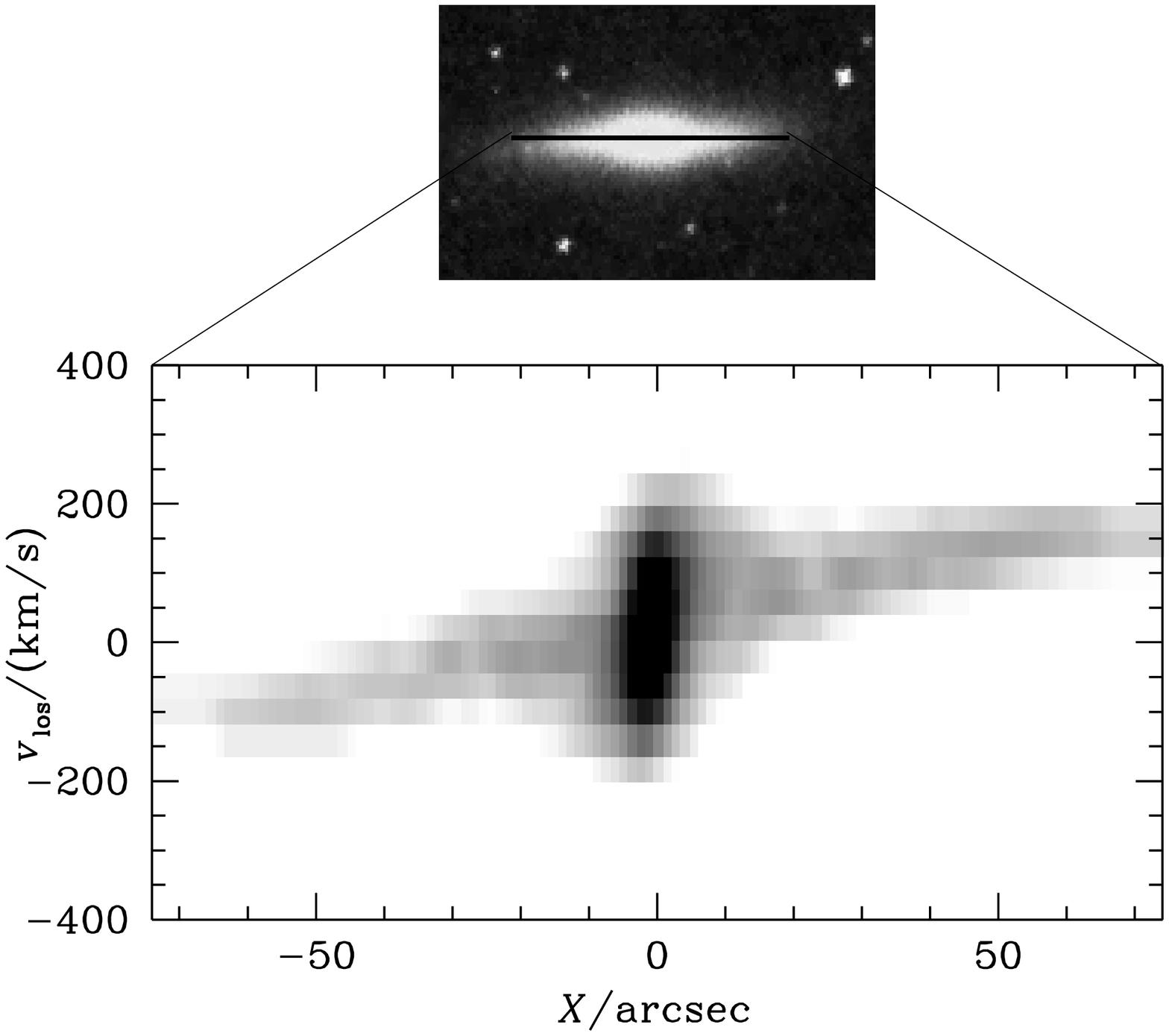}{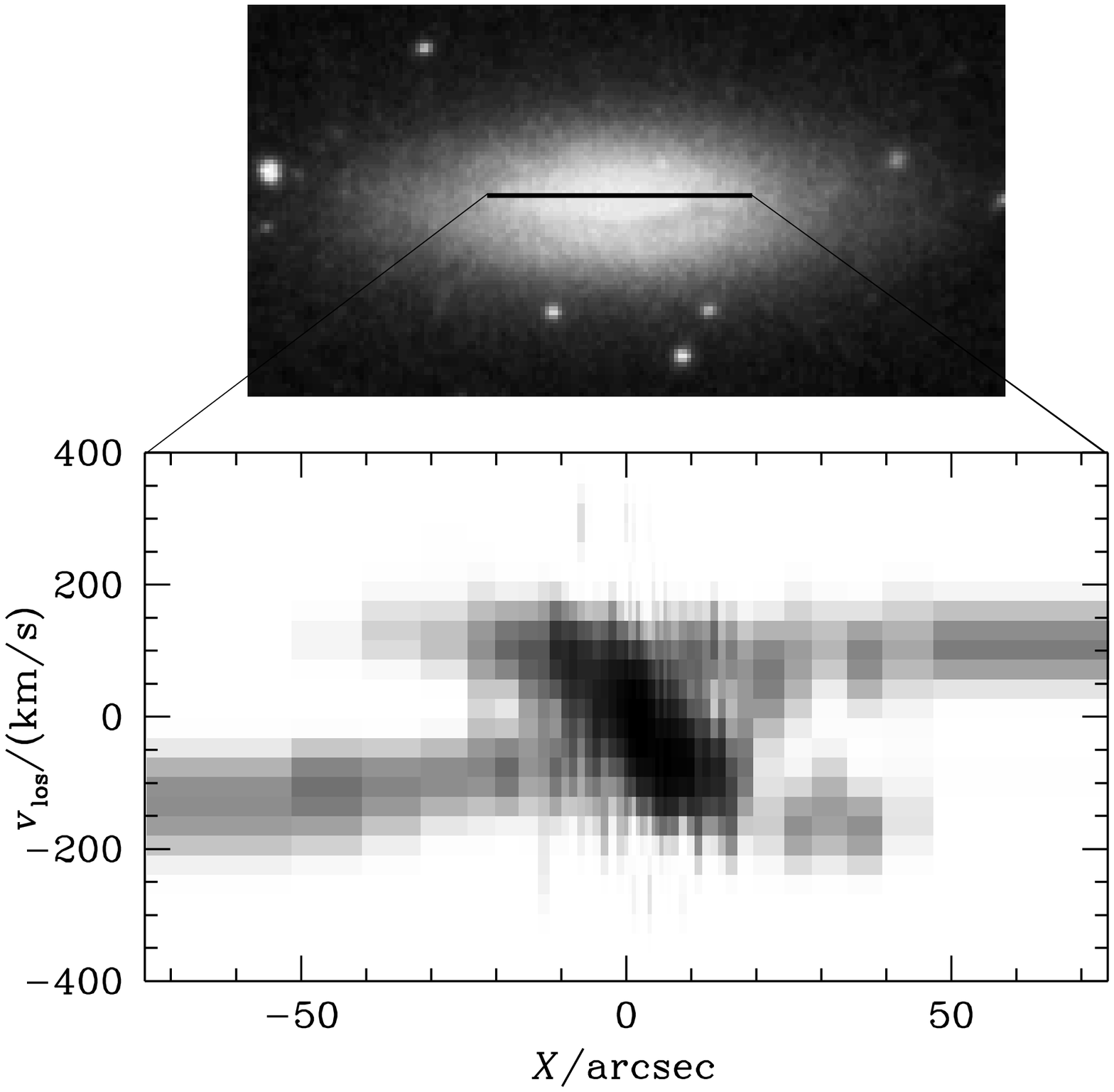}
\caption{Kinematics along the major axes of edge-on disk galaxies.
The axes along which the spectra were obtained are shown in the upper
images; the kinematics are shown in the lower plots, with the
greyscale showing the projected phase density of stars moving with
different line-of-sight velocities at different points along the
observed axes.  Left Panel: the normal S0 galaxy NGC~7332.  Right
panel: the kinematically-peculiar Sa galaxy NGC~3593. [Adapted from
Binney \& Merrifield (1998).]} \label{CRfig}
\end{figure}

Having presented one stellar-kinematic project that depended solely on
measuring mean streaming motions, and a second that involved measuring
the velocity dispersion, we finally turn to a study that requires the
analysis of the complete line-of-sight velocity distribution.  If one
observes the velocity distribution of stars in an edge-on disk galaxy,
one would expect to see the signature of rotation in the data: the
stars on one side of the galaxy should be moving towards us (relative
to the systemic velocity of the whole galaxy), while those on the
other side should be moving away.  Close to the center of the galaxy,
the light will be dominated by the randomly-orbiting bulge stars, so
one would expect to see a larger spread in the velocities in this
region, with less sign of rotation.  When the line-of-sight velocity
distributions of stars are derived from high quality spectra of
edge-on galaxies, the vast majority of systems match this description
(Kuijken, Fisher \& Merrifield 1996) -- a typical example is shown in
the left panel of Figure~\ref{CRfig}.  There are, however, a few
exceptions.

The first kinematically-peculiar stellar disk was discovered by Rubin,
Graham \& Kenney (1992).  They obtained spectra of the
apparently-normal edge-on S0 galaxy NGC~4550, and found that at each
point away from the center of the galaxy the spectral lines were split
into two, one redshifted and one blueshifted relative to the systemic
velocity of the galaxy.  The interpretation of these data is that
NGC~4550 contains two almost identical co-spatial stellar disks that
rotate in opposite directions.  Since then, a few further examples of
this strange phenomenon have been uncovered.  Merrifield \& Kuijken
(1994) found that NGC~7217 contains a retrograde component, but in
this case, although the counter-rotating disks are comparable in
extent, one contains three times as many stars as the other.
Figure~\ref{CRfig} shows the further example of NGC~3593, discovered
by Bertola et al.\ (1996), where one of the components is dominant at
small radii, but it has a shorter scale-length than the other
component, and therefore becomes the fainter at large radii.  Although
there are now several examples of this phenomenon, it is not common: a
systematic survey of 28 S0 galaxies, using spectral data of sufficient
quality that even 5\% of stars on retrograde orbits would have been
detected, failed to find any systems with counter-rotating components
(Kuijken, Fisher \& Merrifield 1996).

Even though these systems are rare, the existence of a few means that
there must be some mechanism by which they are produced.  One
possibility might be that they form from mergers between disk galaxies
that spin in opposite directions.  However, simulations show that
stellar disks are rather fragile objects, and colliding them
invariably destroys the flattened structure, producing an elliptical
system (Toomre \& Toomre 1972).  Disk-like counter-rotating systems
must be formed by some more gentle process.  One plausible scenario is
as follows.  A galaxy forms by the conventional mechanism of the
gravitational collapse of primordial gas, with a disk forming due to
the angular momentum of the infalling material.  Over time, the gas
disk form stars, creating a conventional galaxy.  Primordial gas will
continue to rain down on the galaxy, and over time the angular
momentum of this infalling material will change.  If the angular
momentum changes slowly, the orientation of the galaxy will adjust to
incorporate the new material.  If, however, the angular momentum
changes rapidly through close to 180 degrees, the new infalling will
collide with any gas left from the initial formation, and the angular
momenta of these components will cancel, dumping material toward the
center of the galaxy.\footnote{In this regard, it is interesting that all
of the known counter-rotating stellar disks are in early-type galaxies
with large bulges -- perhaps these bulges owe their stature to the
large amount of material dumped on them during the formation process.}
Once all the first-generation gas has been swept away, subsequent
infalling material will be able to create a new gas disk that rotates
in the opposite direction to the existing stellar component.  Quite a
few such disk systems, where the gas and stars orbit in opposite
directions, have been documented (Bertola, Buson \& Zeilinger 1992).
Ultimately, the new gas disk may start to produce stars, resulting in
a counter-rotating stellar disk system.

\section{Conclusions} \label{concsec}

Dynamical studies of nearby galaxies have a vital role to play in
studying the formation and evolution of these systems.  First of all,
a galaxy is fundamentally a dynamical entity, so no description of it
can be complete without detailed kinematic information.  A simulation
based on a particular cosmological model may predict the formation of
galaxies that look just like those in the real Universe, but if the
orbits of the material that makes up these simulated systems do not
match those of real galaxies, then the model cannot be correct.
Second, once a star is settled on a particular orbit, it is quite hard
to shift it elsewhere, so the the dynamical ``memories'' of these
systems can be comparable to the age of the galaxy.  Thus, by studying
stellar kinematics, one is tapping into the archaeological record of a
galaxy's formation.

The signature of galactic evolution in the stellar kinematics of a
galaxy can be quite subtle.  Until relatively recently, the quality of
data and the sophistication of analysis techniques were not sufficient
to unearth these clues.  However, as we hope the above examples have
illustrated, the quality of kinematic data now available for stellar
disks means that it has become possible to learn a good deal about all
aspects of the dynamics of these systems.

We have really only just begun to exploit what is literally a new
dimension in the study of galactic structure.  In the near future,
pilot studies like those described above should lead into large-scale
systematic surveys of the stellar-kinematic properties of disk
galaxies, providing us with the insights we need to understand the
processes by which these systems form and evolve.

% Finally, we have a little acknowledgements section.

\acknowledgments

MRM is currently supported by a PPARC Advanced Fellowship (B/94/AF/1840).

% That's the end of the main body of the paper.  Now we will have some
% back matter.

\end{document}